\documentclass{elsart3p}        
\usepackage{graphicx}
\begin{document}
\begin{frontmatter}
\title{Scanning microscopies of superconductors at very low temperatures}
\author[address1]{V. Crespo, A. Maldonado, J.A. Galvis, P. Kulkarni,}
\author[address1,address2]{I. Guillamon,}
\author[address1]{J.G. Rodrigo, H. Suderow\thanksref{thank1}, S. Vieira,}
\author[address3]{S. Banerjee,}
\author[address4]{P. Rodiere}
\address[address1]{
Laboratorio de Bajas Temperaturas, Departamento de F\'isica de la
Materia Condensada, Instituto de Ciencia de Materiales Nicol\'as
Cabrera, Universidad Aut\'onoma de Madrid, E-28049 Madrid, Spain}
\address[address2]{H.H. Wills Physics Laboratory, University of Bristol, Tyndall Avenue, Bristol BS8 1TL, UK}
\address[address3]{Department of Physics, Indian Institute of Technology, Kanpur 208016, U.P., India}
\address[address4]{Institut Neel, CNRS/UJF, 25 Avenue des Martyrs, B.P. 166, 38042 Grenoble Cedex 9, France}
\thanks[thank1]{
* Corresponding author: hermann.suderow@uam.es}
\bigskip
\begin{abstract}
We discuss basics of scanning tunneling microscopy and spectroscopy (STM/S) of the superconducting state with normal and superconducting tips. We present a new method to measure the local variations in the Andreev reflection amplitude between a superconducting tip and the sample. This method is termed Scanning Andreev Reflection Spectroscopy (SAS). We also briefly discuss vortex imaging with STM/S under an applied current through the sample, and show the vortex lattice as a function of the angle between the magnetic field and sample's surface.
\end{abstract}
\end{frontmatter}

\section {Introduction}

Using Scanning Tunneling Microscopy and Spectroscopy (STM/S), one can, in principle, image many features of superconductors. In particular, the position of vortices and the symmetry of the lattice, as well as the spectroscopy inside and around vortex cores\cite{Fischer07,Rodrigo04b}. A vortex consists of supercurrents surrounding a magnetic core, and the Cooper pair wavefunction slips its phase by 2$\pi$ in a ring around its center \cite{A57,Blatter94,Brandt95}. The quantized finite circulation implies that the supercurrent velocity has to diverge at the center of the vortex. This divergence is avoided as the Cooper pair wavefunction drops continuously towards the center and becomes zero exactly at the vortex core. Materials Science and nanostructuring have been used to highlight different aspects of vortices, giving new physics or applications. For example, vortex lattice elasticity studies treat vortices as ropes with a rigid center in 3D or as disks in 2D, and are important in explaining vortex lattice disorder and melting \cite{BookVictor,Giamarchi02,Banerjee04,Guillamon09Nat}. In anisotropic superconductors, non-local physics leads to anomalous vortex lattice symmetries\cite{Sakata00,Eskildsen01}. Magnetism may increase vortex density or change its symmetry\cite{Eskildsen98,Guillamon10}. Conductance or T$_c$ oscillations, observed in regular arrays of dots or other nanostructures, highlight vortex pinning in terms of force balance\cite{Silhanek10}. The observation of peaks in the density of states in the center of the vortex cores is understood in terms of the smooth changes of the Cooper pair wavefunction when approaching the center of the vortex\cite{Hess90,Hayashi98,Nishimori04,Guillamon08}.

In this paper we review some aspects of tunneling spectroscopy relevant for the study of superconductors, and discuss recent developments enabling new imaging possibilities, such as Scanning Andreev Reflection Spectroscopy. Measurements have been taken in several dilution and $^3$He refrigerator STM/S set-ups described elsewhere\cite{Rodrigo04b,Suderow11,Maldonado11}.

\begin{figure}[hbt]
\begin{center}\leavevmode
\includegraphics[width=7.2 cm]{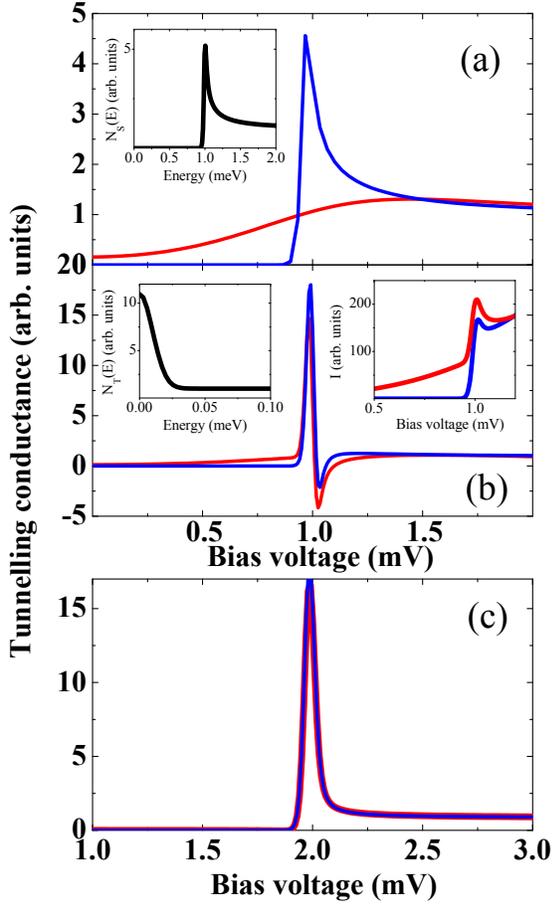}
\vskip -0 cm
\caption{a) Calculated tunneling conductance between a normal tip and a s-wave BCS superconductor as a function of the bias voltage at T$_c$/65 (blue line) and at T$_c$/2 (red line). We use N$_S$(E) as shown in the inset (with peak at $\Delta$=1.76$k_BT_c$=1meV) . b) Tunneling conductance between a tip having a Gaussian shaped peak of the density of states at the Fermi level, with a width of 15$\mu$eV (N$_T$(E) as shown in the left inset) and a superconducting sample with the same density of states N$_S$(E) as in a, at T$_c$/65 and at T$_c$/2. Inset shows the corresponding I(V) curves. c) Tunneling conductance between tip and sample with the same density of states as in a) at T$_c$/65 and at T$_c$/2. Bias voltage labels for c are given in mV at the bottom, and for a and b in mV at the top.}\label{Fig1}\end{center}\end{figure}

\section{Spectroscopy with superconducting tips}

One objective of a tunneling spectroscopy imaging experiment in a superconductor is to obtain with highest precision the energy dependence of the density of states of the superconductor $N_S(E)$ as a function of the position at all temperatures or magnetic fields. The tunneling current $I$ between tip and sample can be written as  $I(V) \propto \int dE [f(E-eV)-f(E)] N_T(E-eV) N_S(E)$, and is the result of the convolution of the densities of states of tip $N_T$ and sample $N_S$, weighted by the Fermi function $f(E)$\cite{Fischer07,Berthod11}. When the tip is made of a non superconducting metal, $N_T$ is energy-independent, and the differential tunneling conductance $\sigma=dI/dV$ is simply $N_S(E)$ smeared by the Fermi function\cite{Fischer07,Song10}. When $k_BT$ is low enough, $\sigma$ is practically equal to $N_S(E)$, but when increasing the temperature, the tunneling conductance may be significantly smeared (Fig.1 a), and measurement uncertainty coming from noise significantly influences the final de-convolution\cite{Crespo06a}.

Consider now, for example, a tip with a $\delta$ function like sharp peak at the Fermi level, in an otherwise finite and energy independent density of states, such as the one produced by a localized state close to the tip\cite{Balatsky06,Ternes09}. In that case, the I(V) curve directly reflects the density of states of the sample, with a peak at the gap edge. The conductance $\sigma$ has correspondingly pronounced maxima and minima, and passes through zero exactly at the position of the peak in I(V). When increasing temperature, the current below the gap edge increases exponentially, reflecting the thermal behavior of the tail of the Fermi function\cite{Maldonado10}, but features in $\sigma$ at the gap edge are largely unaffected (Fig.\ref{Fig1} b). If the density of states of the tip is well known and independent of temperature, then the de-convolultion of the sample's density of states from the I-V curve or from the tunneling conductance remains precise when increasing temperature. Localized state features which give peak shaped densities of states have been observed in some cases close to defects or impurities\cite{Balatsky06,Ternes09,Schmidt10,Aynajian10}. Tips with impurities located close to the apex could therefore give interesting results, although they still need to be realized in a controlled way.

A superconducting tip leads to even better results. The form and properties of the density of states of tips made of well known superconductors have been previously reported with detail\cite{Rodrigo04b,Rodrigo03,Rodrigo04,Guillamon07,Suderow09njp,Stip1,Stip2,Bergeal06,Noat10,Suderow02}. The quasiparticle peaks located at the gap edge lead to a sharp anomaly at the sum of the tip and sample's superconducting gaps (Fig.1c). When increasing temperature, the width of the corresponding anomaly does not change, making the determination of the LDOS of the sample through de-convolution accurate at all temperatures \cite{Rodrigo04c,Crespo09}.

\begin{figure}[hbt]
\begin{center}\leavevmode
\includegraphics[width=6.5 cm]{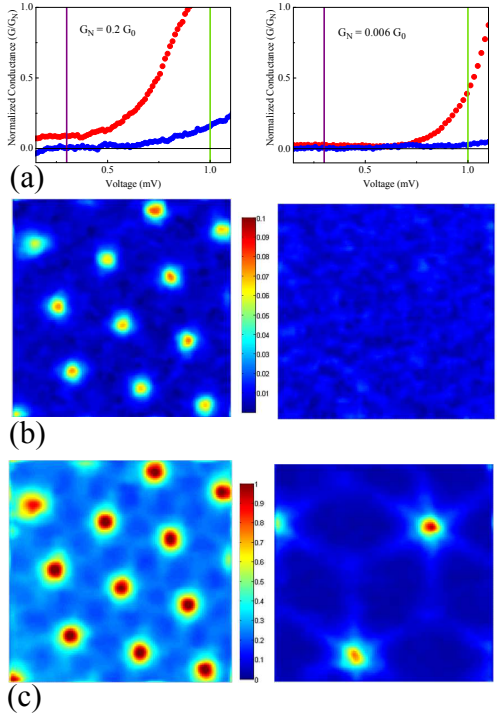}
\vskip -0 cm
\caption{Scanning Andreev Spectroscopy of the vortex lattice of NbSe$_2$ using a tip of Pb at 0.3 K. In left panels we show experiments on the vortex lattice with a short tip-sample distance (conductance at 4 mV is of 17$\mu$S, Andreev reflection is observed), and in the right panels we show experiments made using a larger tip-sample distance (conductance at 4 mV is of 0.5$\mu$S, Andreev reflection is suppressed due to the high tunneling barrier). In a) we show the tunneling conductance at a vortex core (red points), and in between vortices (blue points). Higher conductance at the vortex cores is due to Andreev reflection between the superconducting tip and the NbSe$_2$ sample, only visible when tip-sample distance is short enough (here at 17$\mu$S). Violet and green lines show the voltages at which the spectroscopic images in b anc c are built of. The images built with the tunneling conductance at 0.5 mV (shown in b in violet frames), clearly show the vortex lattice if the tip-sample distance is short (left panel, 17$\mu$S) due to Andreev reflection between the superconducting tip and the normal vortex core. The image built with the tunneling conductance at the same voltage is featureless when the tunneling conductance is low (right panel, 0.5$\mu$S). Color code is shown as a bar at the center using the same scale as in the y axis of a, namely the tunneling conductance normalized to the value at high voltages. In c we show the images constructed from the data at 1 meV, using the color code from the bar at the center. Observed features (star shaped vortex core) do not depend on the tip-sample distance, and have been discussed in previous publications\protect\cite{Rodrigo04b,Rodrigo03,Rodrigo04,Guillamon07,Suderow09njp,Stip1,Stip2,Bergeal06,Noat10}.}
\label{Fig5}\end{center}\end{figure}

\section {Scanning Andreev Reflection Spectroscopy with superconducting tips.}

Superconducting tips also allow, on the other hand, to probe the Josephson effect at the local level\cite{Rodrigo04,Bergeal08,WeinBookVictor}, and they could enable designing, in future, local phase sensitive experiments\cite{Rodrigo04b,Hanaguri09}. There is yet another important feature which has not been explored until now. In particular, at the lowest temperatures, the current due to thermal excitations at voltages below the superconducting gap is exponentially small. Andreev reflection between tip and sample is always present, and the corresponding current depends exponentially on the tunneling barrier\cite{A64}. Therefore, reducing the tunnel barrier at the lowest temperatures leads to the observation of a small current at bias voltages below the superconducting gap which exponentially grows when the distance between tip and sample is reduced\cite{BTK82}. For instance, using a superconducting tip and a normal sample, with a tunneling conductance of around some tens of $\mu$S (i.e. some 0.1 G$_0$, where G$_0$ is the quantum of conductance), the Andreev reflection accounts for the observation of a tunneling conductance of some per cent the high voltage conductance. When scanning with a superconducting tip a superconducting sample, Andreev reflection is further suppressed, as the process requires bouncing back and forth four electronic states, instead of the two states involved in a N-S Andreev reflection process\cite{Oetal83}.

Thus, scanning with, e.g. a superconducting tip of Pb, on a sample with superconducting and normal areas, leads to a conductance at a finite voltage below the superconducting gap which changes as a function of the position, being significantly larger on top of the normal areas. Of course, the experiment needs to be made at low enough temperatures to ensure that the result does not come from thermal excitations. Moreover, local variations in the magnetic field may result in changes in the tip's density of states which induce pair breaking effects, producing similarly a higher low bias conductance\cite{Guillamon07}. Changing the tip-sample distance is the most effective way to separate Andreev processes from other variations of the low bias voltage conductance as a function of the position. In Fig.2 we show results taken at a conductance of 17 $\mu$S (left panels of Fig.2), and compare them with results taken at a conductance of 0.5 $\mu$S (right panels of Fig.2), which is around 0.39\% G$_0$. While thermal excitations or pair breaking depend on the actual density of states of the tip and should be present at all tunneling tip-sample distances, Andreev reflections will decrease exponentially with distance. The left panel of Fig.2b shows the vortex lattice of NbSe$_2$ at 0.1 T viewed by plotting as a function of the position the value of the Andreev conductance. The signature of vortices is washed out when images are taken at low tunneling conductance, below $\mu$S, as shown in the right panel of Fig.2b. When the image is built from the value of the conductance at 1 mV the observed features are due to changes in the local density of states, and have been discussed previously\cite{Rodrigo04b,Rodrigo03,Rodrigo04,Guillamon07,Suderow09njp,Stip1,Stip2,Bergeal06,Noat10}. Namely, the density of states at the center of the vortex core is high and leads to high conductance at 1mV when using a superconducting tip of Pb, and the characteristic star-shape found in NbSe$_2$ is observed in-between vortices. This observation is independent of the value of the tunneling conductance, as is shown in Fig.2c.

This demonstrates that the localized electronic states inside the vortex cores can carry an Andreev current, which peaks at the vortex core. Note that Andreev reflection between a conventional s-wave superconductor and a metal with full spin polarization is not allowed. Thus, Scanning Andreev Spectroscopy can be developed in future as an effective tool to study local spin polarization of materials at very low temperatures.

\section {Current drive scanning tunneling microscopy and spectroscopy.}

\begin{figure}[hbtp]
\begin{center}\leavevmode
\includegraphics[width=6 cm]{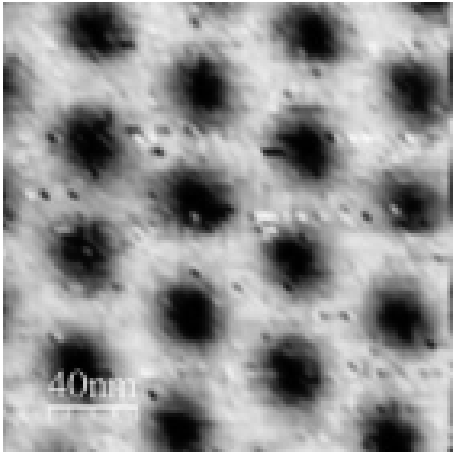}
\vskip -0 cm
\caption{The pinned vortex lattice of NbSe$_2$ observed using conventional scanning tunneling spectroscopy with a normal tip under an applied current of 10 mA at a magnetic field of 0.5 T and at 4.2 K. To obtain the image, full I-V curves are taken at each position. The conductance at zero bias is plotted as a function of the position. At this temperature (4.2 K), the dependence of the I-V curves within vortex cores and in between vortices as a function of the applied current is mixed with the temperature smearing.}\label{Fig3}\end{center}\end{figure}

It is also important to image varying relevant thermodynamic and transport parameters, such as an applied current. Several STM/S experiments have been designed for topography measurements under an applied current. The arrangement requires to ground one side of the sample and apply a bias to the other side, with a current fixed by the resistance of the sample and the bias voltage\cite{Maldonado11,Kirtley88}. In order to study a superconductor, one requires knowledge of the full $N_S(E)$ dependence. Recently, we have developed new STM/S circuitry and methods to be able to make I-V measurements as a function of the position and the current using normal tips\cite{Maldonado11}. In Fig.3 we show a vortex lattice image of NbSe$_2$ obtained by measuring at each position a full I-V curve, and applying a current of 10 mA through the sample, at 4.2 K. Corresponding current density is of 0.4 10$^4$A/m$^2$. Current flow can be expected to be homogeneous, because the London penetration depth $\lambda$ is much larger than the coherence length $\xi$\cite{Guillamon08}. Here we plot the normalized zero voltage conductance as a function of the position\cite{Rodrigo04b}. The vortex lattice is clearly viewed, so that the current is below the de-pinning value. We image therefore the pinned vortex lattice. Further measurements at lower temperatures, and as a function of the current and the magnetic field are under way.

\section {Scanning tunneling microscopy and spectroscopy in a vector magnetic field.}

\begin{figure}[hbtp]
\begin{center}\leavevmode
\includegraphics[width=7.5 cm]{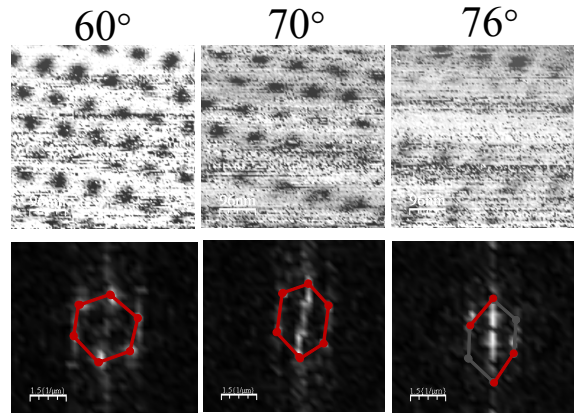}
\vskip -0 cm
\caption{Vortex lattice of NbSe$_2$ as a function of the angle of the applied magnetic field with respect to the surface. Images are taken at 0.5 T and 1.2K. Top panels show the real space images, and bottom panels their Fourier transforms. Red lines and points highlight the Bragg peaks of the vortex lattice. When increasing the angle, the hexagon elongates, and leaves practically a stripe like arrangement at 76$^o$. Grey lines and points show the position of Bragg peaks with smaller amplitude.}\label{Fig4}\end{center}\end{figure}

The control over the direction of the magnetic field is another important aspect which allows studying highly anisotropic samples. In particular, the vortex lattice in layered compounds is expected to show many peculiar features related to the modifications of intervortex potential when the magnetic field is turned from being perpendicular to the layers\cite{Brandt93,Hess94,Buzdin09}. Recently, we have developed a dilution refrigerator three-axis vector magnet STM/S set-up, and could measure the angular dependence of the vortex lattice in NbSe$_2$. The density of vortices decreases when the z-axis component of the magnetic field drops, following $\cos(\theta-\theta_m)$, with $\theta_m$ being the misalignment of the normal to the surface of the sample with respect to the z component of the magnetic field (here $\theta_m=4^o$). As observed previously in Ref.\cite{Hess94}, the vortex lattice buckles, losing hexagonal symmetry, because it tries to orient one of the three high symmetry axis of the hexagon along the planar component of the magnetic field vector. Gradually, the hexagon is elongated and a zig-zag structure which ends up in stripe-like vortex arrangements is found. Although the vortices have been observed (Fig.4), more precise and clear measurements are needed to show this interesting feature of the tilted vortex lattice.

\section {Summary and outlook.}

Scanning tunneling spectroscopy has evolved as a very efficient and complete microscopy of the superconducting state. Work in superconducting tips allow to increase considerably the precision in measuring the density of states of the sample, and provides for new modes of operation. Scanning Andreev Spectroscopy (SAS) appears as an efficient way to probe Andreev reflection at the very local level, and should lead to new imaging technique sensitive to the spin polarization of the tunneling current. Further developments in measurements as a function of an applied current (CD/STS), and in a vector magnetic field, will enable new imaging possibilities. The full potential of Materials Science, with anisotropies of single crystalline materials, and of engineered Nanoscience, with phenomena related to vortex flow in different geometries and arrangement of nanostructures in superconducting samples, come into the realm of the imaging possibilities of STM/S. 

\begin{ack}
We specially acknowledge support from NES program. This work was also supported by the Spanish MEC (Consolider Ingenio Molecular Nanoscience CSD2007-00010 and FIS2008-00454 and ACI2009-0905 programs) and by the Comunidad de Madrid through program Nanobiomagnet. The Laboratorio de Bajas Temperaturas is associated to the ICMM of the CSIC.
\end{ack}


\end{document}